\documentclass[twocolumn]{aastex631}

\usepackage{graphicx}
\usepackage{amsmath}
\usepackage{float}
\usepackage{natbib}
\usepackage{tabularx}
\usepackage{bm}
\usepackage{appendix}
\usepackage{longtable}
\usepackage{enumitem}
\usepackage{booktabs}
\usepackage{multirow}

\shorttitle{Luminous CO Binaries from Disrupted Dense Clusters}
\shortauthors{Schiebelbein-Zwack et al.}

\graphicspath{{./}{figures/}}

\begin{document}

\title{The Contribution of Disrupted Dense Star Clusters to Gaia’s Compact Object Binaries}

\correspondingauthor{Aryanna Schiebelbein-Zwack}
\email{aryanna.schiebelbein@mail.utoronto.ca}

\author[0009-0003-3908-6112]{Aryanna Schiebelbein-Zwack}
\affiliation{David A. Dunlap Department of Astronomy and Astrophysics, University of Toronto, 50 St George St, Toronto, Ontario M5S 3H4, Canada}
\affiliation{Canadian Institute for Theoretical Astrophysics, University of Toronto, 60 St. George Street, Toronto, Ontario M5S 3H8, Canada}

\author[0000-0001-9582-881X]{Claire S.\ Ye}
\affiliation{Canadian Institute for Theoretical Astrophysics, University of Toronto, 60 St. George Street, Toronto, Ontario M5S 3H8, Canada}

\author[0000-0002-8556-4280]{Marta Reina-Campos}
\affiliation{Canadian Institute for Theoretical Astrophysics, University of Toronto, 60 St. George Street, Toronto, Ontario M5S 3H8, Canada}
\affiliation{Department of Physics \& Astronomy, McMaster University, 1280 Main Street West, Hamilton, L8S 4M1, Canada}
\affiliation{Instituto Galego de Física de Altas Enerxías, Universidade de Santiago de Compostela, 15782 Santiago de Compostela, Galicia, Spain}

\author{Aeysha Munawwarah}
\affiliation{David A. Dunlap Department of Astronomy and Astrophysics, University of Toronto, 50 St George St, Toronto, Ontario M5S 3H4, Canada}

\author[0000-0002-6871-1752]{Kareem El-Badry}
\affiliation{Department of Astronomy, California Institute of Technology, 1200 E. California Blvd, Pasadena, CA, 91125, USA}

\author[0000-0002-1386-0603]{Pranav Nagarajan}
\affiliation{Department of Astronomy, California Institute of Technology, 1200 E. California Blvd, Pasadena, CA, 91125, USA}

\begin{abstract}
We present the first model of the Milky Way’s detectable compact object--luminous star binary population from disrupted dense star clusters. We bridge large-scale cosmological star cluster formation with high-resolution dynamical evolution of compact object binaries by mapping the predicted star clusters from the EMP-\textit{Pathfinder} simulations to $N$-body \texttt{Cluster Monte Carlo} models. We predict that approximately $3\times10^5$ white dwarfs (WDs), $1.5\times10^5$ black holes (BHs), and $1\times10^3$ neutron stars (NSs) in binaries with luminous companions are released to the Galaxy from now-disrupted dense star clusters throughout the history of the Milky Way. Synthetic observations modeled with the \texttt{gaiamock} pipeline reveal that the modeled Gaia DR3 yields are sparse ($\approx 2$ WDs, 0 NS, 0 BHs at 90\% credibility), with the majority lying beyond the detection horizon. Gaia DR4 is expected to increase the observational yield of these systems only marginally, as the benefits of an expanded search volume are largely offset by the diminished astrometric and photometric precision of more distant sources ($\approx 14$ WDs, 0 NS, 0 BHs). While the underlying BH binary population is similar to that of WDs, they are detected far less frequently; they tend to pair with lower-mass, dimmer companions and have less temporal coverage of their long orbital periods. For NSs, we suggest that the observed over-representation of metal-poor, halo systems is inconsistent with an origin in disrupted dense star clusters. Instead, the observed Gaia NS population could reflect the accretion history of metal-poor, dwarf galaxies into the Milky Way, isolated binary star evolution, or supernova physics.
\end{abstract}

\section{Introduction} \label{sec:intro}

The Gaia survey mission has collected astrometry for over a billion stars, $\approx 1\%$ of the Galactic population \citep{2016A&A...595A...1G, 2021A&A...649A...1G, 2024NewAR..9801694E}. Gaia's third data release (DR3) contains the first orbital solutions for their binaries, expanding the sample in the literature by a factor of 50 \citep{2021A&A...649A...1G, 2024NewAR..9801694E}. This wealth of data has transformed the landscape of stellar binary studies with expanded coverage of binary parameters and increased discovery potential for exotic and rare configurations informing models of binary star formation and evolution \citep[e.g.][]{Chawla+2022,2025A&A...704A...6R,Yamaguchi+2025}. Furthermore, these binaries' Galactic kinematics and orbital architectures may also be influenced by the dynamics of their host star clusters \citep[e.g.][]{Kremer+2020catalog, 2024ApJ...977..203C}, the Galactic tidal field \citep[e.g.][]{2014ApJ...782...60K}, and the Milky Way assembly history \citep[e.g.][]{2019MNRAS.482L.139E, 2020ApJS..246....4T, 2024MNRAS.535..949B}. 

Following Gaia's DR3 \citep{2021A&A...649A...1G}, observational catalogs of candidate white dwarfs \citep[WDs; e.g.,][]{2021MNRAS.506.2269E, 2023MNRAS.518.2991S}, neutron stars \citep[NSs;][]{2024OJAp....7E..58E}, and black holes \citep[BHs;][]{El-Badry+2023_bh1, El-Badry+2023_bh2, 2024A&A...686L...2G} as unseen companions to luminous stars are rapidly emerging. Among the most striking discoveries is Gaia BH3, a $\approx 33 M_\odot$ BH with a metal-poor ([Fe/H]$\approx -2.6$) companion \citep{2024A&A...686L...2G,  2024OJAp....7E..38E, 2024A&A...688L...2M}. This system has been chemically and kinematically associated with the ED-2 stellar stream, originating from a disrupted star cluster \citep{2024A&A...687L...3B, 2024OJAp....7E..38E, 2024A&A...688L...2M}. Similarly, there are potential hints of cluster origins for the Gaia NS systems. Observed NS binaries with metal-poor companions and halo orbits are significantly over-represented: they constitute $\approx 14\%$ (3 out of 21) of candidate NS systems, in stark contrast to the $\sim 0.5\%$ (61 out of 11,420) found among binaries in the parent Gaia sample \citep{2024OJAp....7E..58E}. Discoveries such as these are only expected to increase with future data releases; Gaia DR3 produced $\approx 170,000$ orbital solutions, and DR4 is expected to increase this by at least a factor of $\approx 5$ \citep{2024OJAp....7E.100E}. 

In addition to being frequently located in the halo, these Gaia {compact object} (CO) binaries are also notable for their typically wide, eccentric orbits. The NS candidates have a median eccentricity of $e=0.4$ and periods ranging from $100-1000$ days \citep{2024OJAp....7E..58E}. Similarly, the Gaia BHs have $e>0.4$ and periods ranging from $10-1000$s of days \citep{El-Badry+2023_bh1, El-Badry+2023_bh2, 2024A&A...686L...2G}. While the WDs binaries have a population of low-mass WDs ($M < 0.8 M_\odot$) in low-eccentricity ($e < 0.2$) orbits \citep{2023MNRAS.518.2991S}, a subset of more massive WDs ($M > 1.0 M_\odot$) are in binaries that display a broad eccentricity distribution, with a tail extending to $e \approx 0.8$ \citep{2024OJAp....7E..58E}.

Several studies have sought to explain the origins of these wide, eccentric binaries, yet their stellar phylogenies remain elusive. Isolated binary evolution requires fine-tuning of mass transfer processes and supernova natal kicks. The present-day orbital separations of many of these WD, NS, and BH binaries overlap with the radii of the COs' giant star progenitors, implying they should have experienced common envelope evolution, yet their orbits remain wide and eccentric \citep{2024OJAp....7E..58E, 2024A&A...686L...2G, 2024MNRAS.529.3729S, 2024PASP..136h4202Y, 2025PASP..137j4205Y}. Moreover, systems like Gaia NS1 ($M = 1.90 \pm 0.04\,M_\odot$; $e=0.122 \pm 0.002$) challenge supernova kick prescriptions, as current models often predict stronger natal kicks for more massive NS remnants \citep{2023arXiv231112109B, 2024OJAp....7E..27E}. 

Dynamical formation in star clusters has been invoked to address the challenges of isolated binary evolution and explain Gaia BH3 in the ED-2 stream. Indeed, recent work by \citet{MarinPina+2024} showed that dense star clusters are capable of producing Gaia BH3-like systems. However, the broader results are conflicting. Some studies find that young and open star clusters efficiently produce wide, eccentric BH--star binaries but overpredict the Gaia DR3 observations, while simultaneously finding that isolated binaries cannot reproduce observations \citep{DiCarlo+2024,2025PASP..137d4202N}. Others suggest that the formation efficiency of detached BH--star binaries in young and open clusters is comparable to isolated binary evolution \citep{Kotko+2024}. 

Overall, while these studies have consistently suggested that star clusters can produce Gaia BH binaries, the formation of Gaia NS binaries remains underexplored. One study, \citet{Tanikawa+2024}, argues that these systems are unlikely to originate from open star clusters because most NSs are ejected via supernova natal kicks. Instead, they propose that these systems originate from isolated binaries, which requires the aforementioned fine-tuning. 

Given the aforementioned association of Gaia BH3 with a now-disrupted star cluster and the overabundance of metal-poor Gaia NSs on halo orbits, disrupted dense clusters emerge as a potentially rich yet largely unexplored channel for binary star production. Since most clusters that formed at high redshift, similar to today’s metal-poor GCs in the Milky Way, have since dissolved into the Galactic field, the Gaia binaries found in the field today could potentially be surviving remnants of these dynamical environments. The contribution of disrupted dense star clusters to various types of Gaia CO systems remains to be determined.

Inspired by these recent perplexing discoveries and the discovery potential of upcoming data releases, we explore for the first time the impact of dissolved star clusters on the Gaia-like CO population in the Milky Way. In this work, we bridge the gap between the small-scale physics of binary star evolution and stellar dynamics in dense star clusters with the large-scale evolution of star clusters within the Galaxy. Our approach differs from previous work in two major aspects. Firstly, we focus on disrupted dense clusters, using the properties of star clusters from the EMP-\textit{Pathfinder} simulations, a state-of-the-art suite of hydrodynamic galaxy formation simulations with sub-grid star cluster evolution and disruption \citep{Reina-Campos+2022}. Secondly, this work expands the analysis to all types of stellar remnants. By combining the EMP-\textit{Pathfinder} clusters with the stellar dynamics modeled by the \texttt{Cluster Monte Carlo} (\texttt{CMC}) code \citep{CMC1,Kremer+2020catalog}, we track the journey of CO binaries from their birth in dense environments to their eventual release and evolution in the Galactic field. 

This paper is organized as follows. In Section \ref{sec:emp}, we describe the EMP-\textit{Pathfinder} cosmological simulations, which produce the first key ingredient in our model, the dense star clusters. In particular, we discuss how cluster evolution and dissolution are modeled, and examine the properties of the simulated star clusters. We describe the \texttt{CMC} models in Section \ref{sec:cmc}, which provide our second model ingredient, the internal dynamical processes of star clusters. Our methodology for combining these two models is detailed in Section \ref{sec:COs}, as well as the resulting binaries that emerge from this model. Section \ref{sec:detect} covers the simulation of the Gaia observation pipeline, wherein we determine which of our simulated binaries are observable. Section \ref{sec:discuss} gives an overview of some of the uncertainties and caveats associated with our model, as well as a discussion on the possible formation pathways for the Gaia COs. We conclude in Section \ref{sec:conclu} and give an outlook of what our results indicate for the Gaia CO binary population.

\section{EMP-\textit{Pathfinder}: Star Cluster Formation and Disruption in Cosmological Simulations}\label{sec:emp}

\subsection{ The EMP-Pathfinder Simulations}

Star clusters are inextricably linked to the galaxy they reside within. As a consequence, it is necessary to self-consistently model star clusters alongside their host galaxy to accurately predict their properties and evolution. To do so, we use the simulated star clusters from the EMP-\textit{Pathfinder} simulations \citep{Reina-Campos+2022}. These simulations model the formation, evolution, and disruption of star clusters within galaxies over the lifetime of the Universe using a sub-grid approach\footnote{The simulations are an expanded and upgraded version of the MOdelling Star cluster population Assembly In Cosmological Simulations \citep[MOSAICS;][]{kruijssen11} and the MOSAICS within EAGLE \citep[E-MOSAICS;][]{pfeffer18,kruijssen19} projects.}. In this setup, star clusters form and evolve at run-time as their host galaxies assemble over cosmic time. Of particular interest for this project, the EMP-\textit{Pathfinder} simulations are a suite of zoom-in cosmological simulations of Milky Way-mass galaxies. 

Among the characteristics of the EMP galaxy formation model, one stands out: the inclusion of the physics of the multiphase nature of the interstellar medium is critical for modelling the disruption of star clusters. EMP couples the non-equilibrium heating and cooling networks from the GRACKLE library \citep{smith17} into AREPO \citep{springel10}, which allows the gas to reach temperatures as low as 10 K \citep{Reina-Campos+2022}. By allowing the gas to cool down, the interstellar medium develops the cold, dense clouds that dominate most of cluster disruption via tidal shocks \citep[e.g.][]{baumgardt03,lamers05a,gieles06,elmegreen10,elmegreen10b,kruijssen12,miholics17,Reina-Campos+2022}. Only under these conditions, the simulated old star clusters show a mass distribution in agreement with that of GCs in the Milky Way and M31. 

The fiducial cluster subgrid model makes three important assumptions regarding how star clusters form. The first is that the fraction of the stellar mass forming in gravitationally bound clusters is environmentally dependent, increasing the bound fraction when the interstellar medium has a higher gas pressure \citep{kruijssen12b}. The second assumption is that the initial mass function follows a doubly-exponential Schechter function, where both extremes increase in environments with higher gas pressures \citep{trujillo-gomez19}. Lastly, the initial half-mass radius of the star clusters is set to $r_{\rm h} = 4~{\rm pc}$ and it does not evolve over time. Additionally, the star formation efficiency is constant.

\begin{figure*}
    \centering
    \includegraphics[width=1\linewidth]{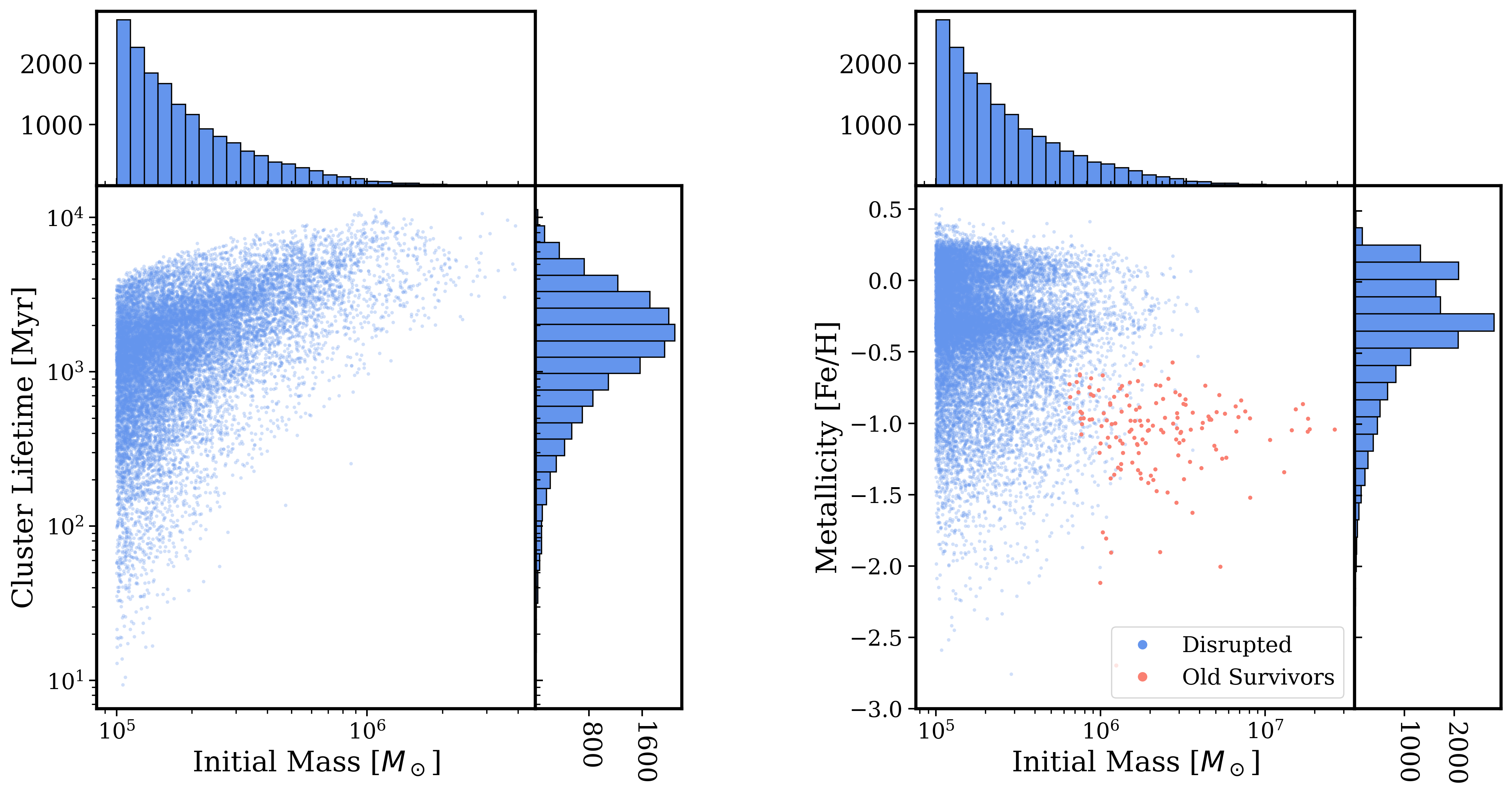}
    \caption{\textit{Left:} The lifetime of EMP-\textit{Pathfinder} disrupted clusters vs. their initial mass. Due to the assumption of a constant initial half-mass radius, these quantities are correlated; more massive clusters tend to have longer lifetimes. \textit{Right:} Cluster metallicity vs. initial mass, for EMP-\textit{Pathfinder} clusters that disrupt (blue), as well as globular cluster analogues (orange). In the simulation, $\approx 70$\% of the clusters disrupt. Of the survivors, only $\approx 2$\% are older than $10$~Gyr and are similar in number and mass to the observed Milky Way GCs \citep{Reina-Campos+2022}. These clusters cover a wide range of metallicities and masses, notably overlapping with the \texttt{CMC} cluster models described in Section \ref{sec:cmc}.}
    \label{fig:emosaic_disruptiontime_mass}
\end{figure*}

As clusters age, they lose mass via three mechanisms: stellar evolution, dynamical processes and dynamical friction. Dynamical processes encompass relaxation and tidal shocks. Two-body interactions within the cluster lead to a slow, but continuous disruption of the cluster \citep[e.g.][]{Ambartsumian38,spitzer40,henon61}, whereas sudden and strong variations of the tidal field (i.e.~tidal shocks) caused by overdensities in the interstellar medium or galactic disk crossings dominate the mass loss of star clusters \citep[e.g.][]{baumgardt03,lamers05a,gieles06,elmegreen10,elmegreen10b,miholics17}. Dynamical friction, whereby clusters lose energy through gravitational interactions with other bodies in the Galaxy and fall towards the Galactic centre, is accounted for in post-processing. Clusters are considered disrupted when the timescale of dynamical friction is shorter than their age. Taking all these processes together, star clusters lose most of their mass via tidal shocks while they orbit their parent galaxy, but shock-driven disruption becomes subdominant if they are ejected to the halo of the galaxy, allowing them to survive for longer.

In addition to the impact of the galactic environment on the formation, evolution, and survival of a cluster, the internal dynamics of the clusters are also of vital importance. The subgrid clusters spatial profile distributions are described by a King profile with concentration parameter $W_0=5$, which is the same initial profile adopted by the \texttt{CMC} models, discussed in Section \ref{sec:cmc}. This ensures that our coupling of these two models is optimal.

\subsection{ Properties of The Disrupted Dense Clusters}

\begin{figure*}
    \centering
    \includegraphics[width=1\linewidth]{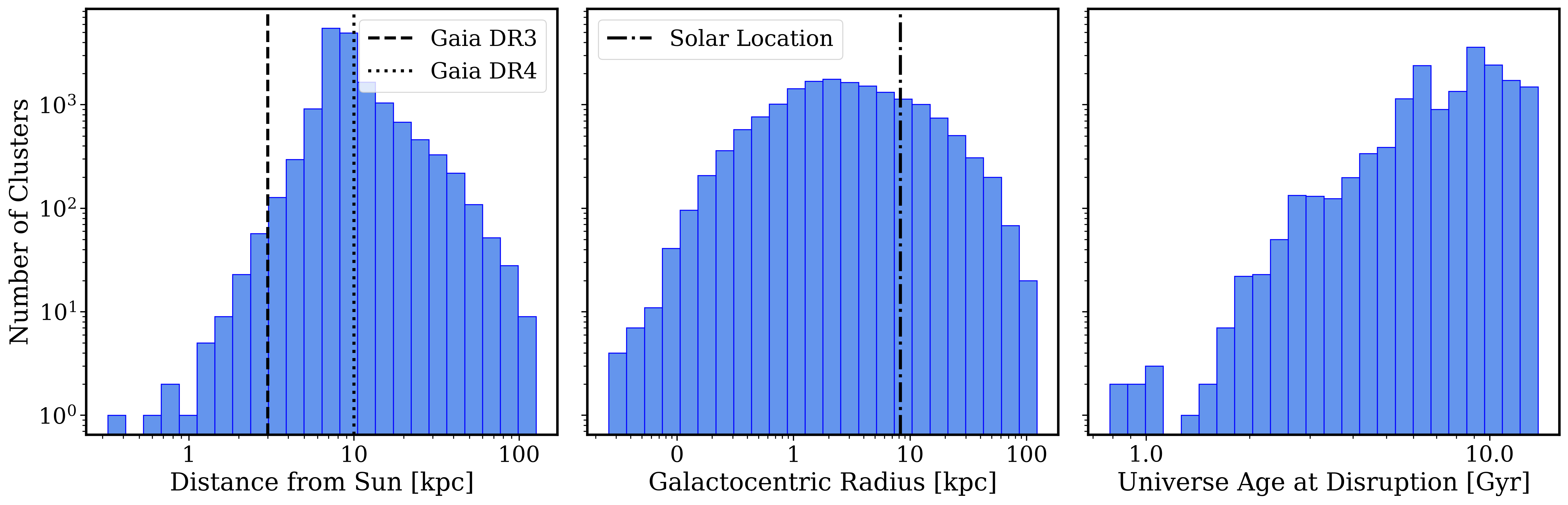}
    \caption{
    \textit{Left:} The cluster distances from the Sun, for a single realization of the Sun's position in the Galaxy. The dashed (dotted) line shows generous upper limits on the detection horizon for Gaia DR3 (DR4). Of the disrupted clusters, $\approx 0.6\%$ are within the DR3 limit of 3 kpc and $\approx70\%$ are within that of DR4 at 10 kpc.
    \textit{Middle:} The galactocentric radii of the clusters in the snapshot directly after disruption. Most clusters disrupt closer to the Galactic centre than the Sun, which is located at 8.24 kpc. These values overlap with those of the \texttt{CMC} models (see Section \ref{sec:cmc}).
    \textit{Right:} The age of the Universe at the time of cluster disruption.}
    \label{fig:emosaic_properties}
\end{figure*}

In this work, we use the simulation \texttt{MW22} from \citet{Reina-Campos+2022}, and focus on studying fully disrupted massive star clusters (i.e.~initially more massive than $10^5~{\rm M}_\odot$ and disrupted by the present day). The properties of the disrupted massive star clusters, similar to GCs and their progenitors ($\geq 10^5\,M_{\odot}$), are shown in Figures \ref{fig:emosaic_disruptiontime_mass} and \ref{fig:emosaic_properties}. As shown in Fig.~\ref{fig:emosaic_disruptiontime_mass}, star clusters with larger initial masses tend to stay bound longer before disrupting due to their deeper gravitational well and longer two-body relaxation time. Since the disruption mechanisms are inversely dependent on the density of the star cluster, the assumption of a constant half-mass radius implies that their mass loss is solely modulated by their mass. Many clusters are disrupted later in their life, as shown by the cluster lifetimes beyond a few Gyr in Fig.~\ref{fig:emosaic_disruptiontime_mass}, and there is adequate time for the dynamical evolution of COs to become relevant. Fig. \ref{fig:emosaic_disruptiontime_mass} also shows the metallicity of the disrupted clusters. This range of metallicities overlaps with those of the \texttt{CMC} models discussed in Section \ref{sec:cmc}. We note that the sample of model clusters from EMP-\textit{Pathfinder} with metallicities near that of Gaia BH3 ([Fe/H]$\approx -2.6$) is sparse. While this highlights the lower metallicity regime as a valuable area for future Galactic simulations to explore, we do not expect this sparsity to bias our results. CO formation appear to be relatively insensitive to metallicity below approximately $0.1\,Z_\odot$ (e.g., $[Fe/H] \lesssim -1$), a range for which our catalog maintains adequate coverage \citep[e.g.][]{2025ApJ...979..209V}.

In Fig. \ref{fig:emosaic_properties}, it can be seen that most disrupted clusters are located beyond the approximate detection horizon of Gaia DR3, which is below 3~kpc \citep{2024NewAR..9801694E, 2024OJAp....7E.100E}. However, the upper distance limit of DR4 \citep[10~kpc;][]{2024OJAp....7E.100E, 2025PASP..137d4202N} coincides with the peak of the disrupted cluster distance distribution from the Sun, as this horizon encompasses the Galactic center where the majority of clusters disrupt. Thus, the discovery potential of binaries originating from disrupted dense clusters will increase immensely in DR4. Note that while some binaries may be detected out to these distance limits, many will remain undetected due to other observational constraints, such as magnitude and orbital period. We account for the combination of these effects on detectability in Section \ref{sec:detect}.

\begin{figure*}
    \centering
    \includegraphics[width=1\linewidth]{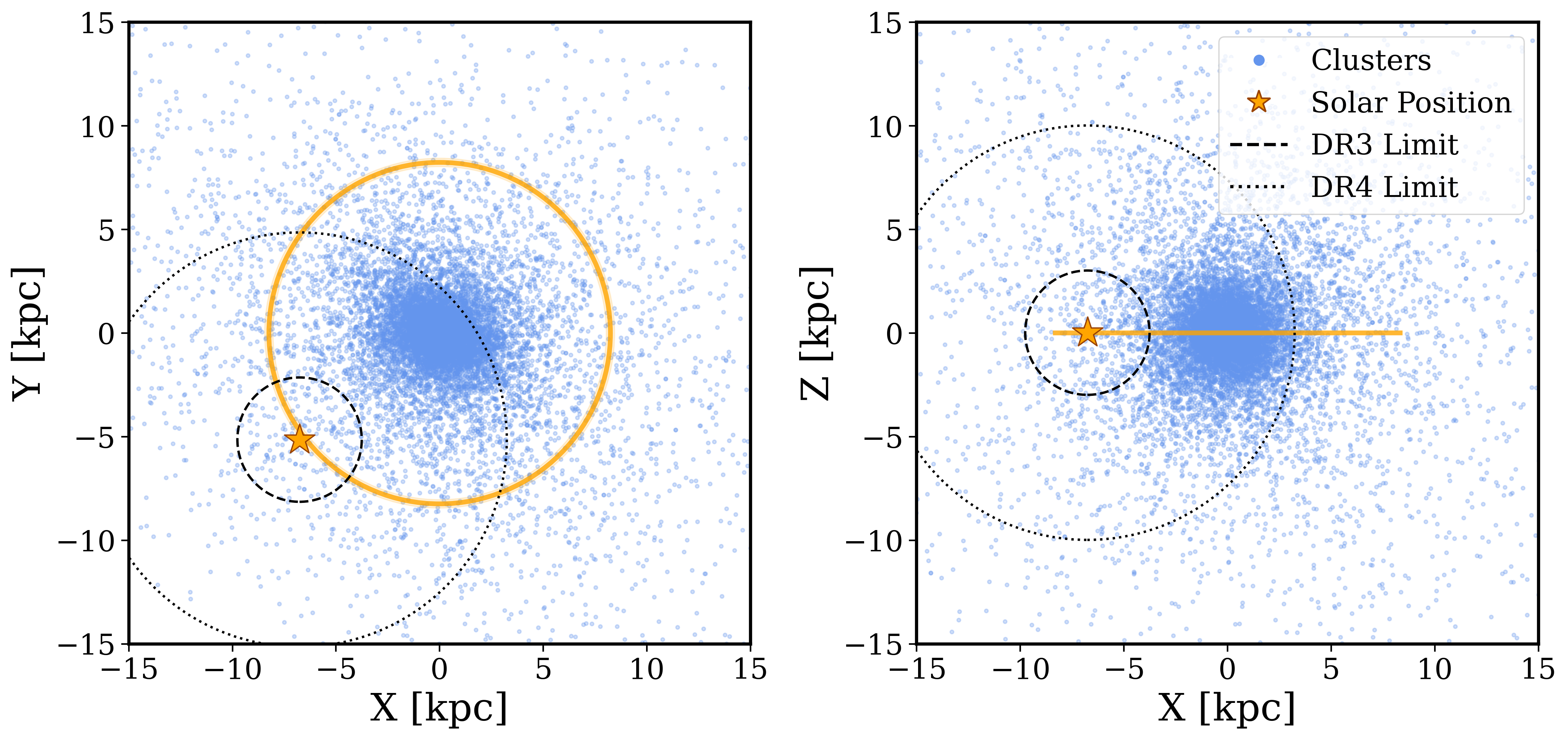}
    \caption{The spatial locations of the disrupted massive star clusters from EMP-\textit{Pathfinder} within a simulated Milky Way-like galaxy, directly following their disruption, in cartesian coordinates. The possible locations of the Sun throughout the Galaxy are shown by the orange annulus. We rotate the location of the Solar System around the X-Y plane to account for the stochastic spatial variation of clusters in the Galaxy relative to an observer. An example realization of the Solar System's location is indicated by the orange star. The Gaia observation horizons are approximate and most binaries will remain undetectable even within the indicated radii.}
    \label{fig:emosaic_coords}
\end{figure*}

The galactocentric radii of the clusters following disruption indicate that most clusters disrupt nearer to the Galactic center than the Sun, which is located at $8.24\pm0.2$ kpc \citep{2023AstL...49..493B}. This is also shown clearly in Fig. \ref{fig:emosaic_coords}, which displays the spatial distribution of the dense clusters throughout the Milky Way. The Sun is shown at $z=0.0208$~kpc and $r=8.24$~kpc \citep{2019MNRAS.482.1417B, 2023AstL...49..493B}. In this work, we compute 100 realizations by rotating the position of the Solar System throughout the Milky Way-mass galaxy. Accounting for the slight variations that occur from rotating the Solar position in the Galaxy, we find that 0.3–0.6\% of the sample lies within 3 kpc of the Sun, while roughly 65–70\% falls within 10 kpc (90\% credible regions). 

We also show the age of the Universe at disruption, which is equal to the sum of the time at cluster formation and the cluster lifetime. The age at disruption determines the length of time over which the binaries are evolved forward to the present day (see Section \ref{sec:COs}).

Overall, the EMP-\textit{Pathfinder} simulations provide a state-of-the-art model of the cluster population within our Galaxy. By incorporating a physically rigorous treatment of cluster disruption, these simulations successfully reproduce the distribution of Milky Way clusters observed today. However, modeling the formation of Gaia binaries requires a detailed treatment of stellar populations and internal cluster dynamics. For this, we turn to $N$-body dynamical models of dense stellar environments.

\section{CMC: Binary Evolution and Dynamics in Dense Star Clusters}\label{sec:cmc}

To characterize CO binaries originating from disrupted dense star clusters, we map the properties of the dissolved clusters from EMP-\textit{Pathfinder} onto Monte Carlo $N$-body simulations run with the \texttt{Cluster Monte Carlo} code \citep[\texttt{CMC};][]{CMC1}. \texttt{CMC} accounts various relevant physics for cluster evolution, including two-body relaxation, strong three- and four-body dynamical interactions, and mass loss through galactic tidal potential. Binary and single star evolution is fully coupled to stellar dynamics and is computed with the binary population synthesis code \texttt{COSMIC} \citep{COSMIC}. We use a catalog of \texttt{CMC} simulations \citep{Kremer+2020catalog} covering a wide range of initial cluster masses ($M=1.2\times10^5$, $2.4\times10^5$, $4.8\times10^5$, and $9.7\times10^5\,M_{\odot}$), initial virial radii ($r_v=$0.5, 1, 2, and 4~pc), metallicities ($Z=$0.0002, 0.002, and 0.02), and galactocentric distances ($r_g=$2, 8, and 20~kpc). These simulations are evolved for a Hubble time, and the model clusters' properties at the present day match well with those of observed Milky Way GCs. These simulations can thus be used to represent dense, massive disrupted clusters.

All \texttt{CMC} catalog models adopt a standard Kroupa initial mass function \citep{Kroupa_2001} and assume a uniform initial binary fraction of $5\%$, independent of the primary mass. The initial properties of the binaries are sampled from a variety of relations. The secondary masses of these binaries are drawn from a uniform mass ratio $q \in [0.1,1]$ of the primary mass \citep{Duquennoy_Mayor_1991}, and the initial binary orbital periods are sampled from a log-uniform distribution independent of mass \citep[e.g.,][]{Duquennoy_Mayor_1991} ranging from near-contact [$\geq 5(R_1 + R_2)$, where $R_1$ and $R_2$ are the radii of the binary component stars] to the hard/soft
boundary. Note that this prevents the model from forming primordial BH binaries with solar-mass companions, as BH progenitors have initial masses $M \gtrapprox 20\,M_\odot$ \citep[e.g.,][]{2023arXiv230409350H}. However, as shown in Fig. \ref{fig:emosaic_disruptiontime_mass}, the average cluster lifetime prior to disruption is several billion years. Over this timescale, most primordial BH binaries likely undergo significant dynamical interactions, effectively shifting their companion mass distribution to match that of the dynamically formed BH binaries in our model. Finally, the initial binary eccentricities are sampled from a thermal distribution \citep[e.g.,][]{Heggie_1975}.

The \texttt{CMC} clusters adopt the `rapid model' for BH and NS formation from core-collapse supernovae \citep{Fryer+2012}. In this model, NSs receive natal kicks $v_{\rm NS}$ drawn from a Maxwellian distribution with velocity dispersion $\sigma_{\rm CCSN} = 265\,{\rm km\,s^{-1}}$ \citep{Hobbs+2005}, while BHs receive fallback-modulated kicks that scale with $v_{\rm NS}$ and the mass fraction $f_{fb}$ of the stellar envelope that falls back during core collapse $v_{\rm BH} = (1-f_{fb})v_{\rm NS}$. In addition, NSs can form via electron-capture supernovae, which occur when the oxygen–neon core of a star or an oxygen–neon WD reaches a critical mass near the Chandrasekhar limit and collapses due to electron capture and the resulting loss of electron pressure \citep[e.g.,][]{Miyaji+1980,Nomoto_1984,Nomoto_1987}. In this case, we assume the NSs receive small natal kicks drawn from a Maxwellian distribution with velocity dispersion $\sigma_{\rm ECSN}=20\,{\rm km\,s^{-1}}$ \citep{Kiel+2008}. Most NSs retained in dense star clusters, where the typical escape velocity is $\lesssim 100\,{\rm km\,s^{-1}}$, form via electron-capture supernovae \citep[e.g., see][for more details]{Ye+2019_msp}.

\section{Compact Object-Star Binaries from Disrupted Dense Star Clusters}\label{sec:COs}

\subsection{Mapping from the EMP-\textit{Pathfinder} Clusters to \texttt{CMC} Binaries}\label{sec:map}

\begin{figure*}[t]
    \centering
    \includegraphics[width=1\linewidth]{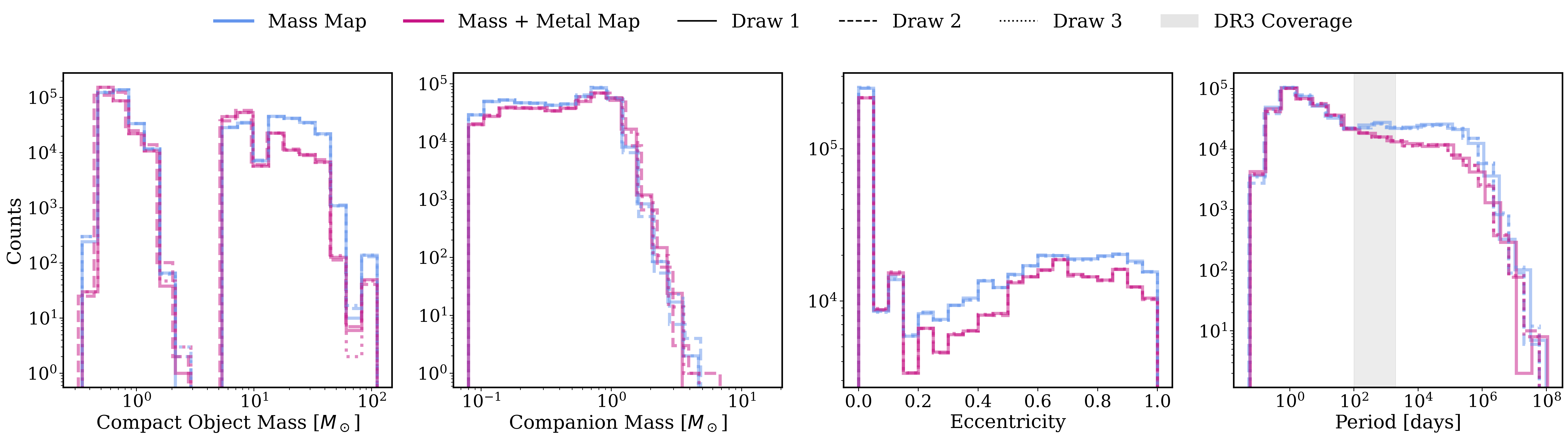}
    \caption{The property distributions of CO binary systems sampled from \texttt{CMC} for different mappings from EMP-\textit{Pathfinder}. Blue lines indicate that the \texttt{CMC}/EMP-\textit{Pathfinder} mapping was done based on mass alone. Pink lines indicate that the mapping additionally accounts for metallicity. The different linestyles represent three different random draws, showing that the results are robust against random noise. The mapping that includes metallicity has slightly less support at higher CO masses because the EMP-\textit{Pathfinder} disrupted clusters have a high concentration at metallicities above $[Fe/H]=-0.3=1/2 Z_\odot$. Since these more massive COs are missing, the `Mass+Metal' map also has fewer binaries with large eccentricities and long periods. An approximate range of observable periods \citep[100-2000 days;][]{2025arXiv251005982M} in Gaia DR3 is shown in shaded grey.}
    \label{fig:histograms_mappings}
\end{figure*}

\begin{figure*}
    \centering
    \includegraphics[width=1\linewidth]{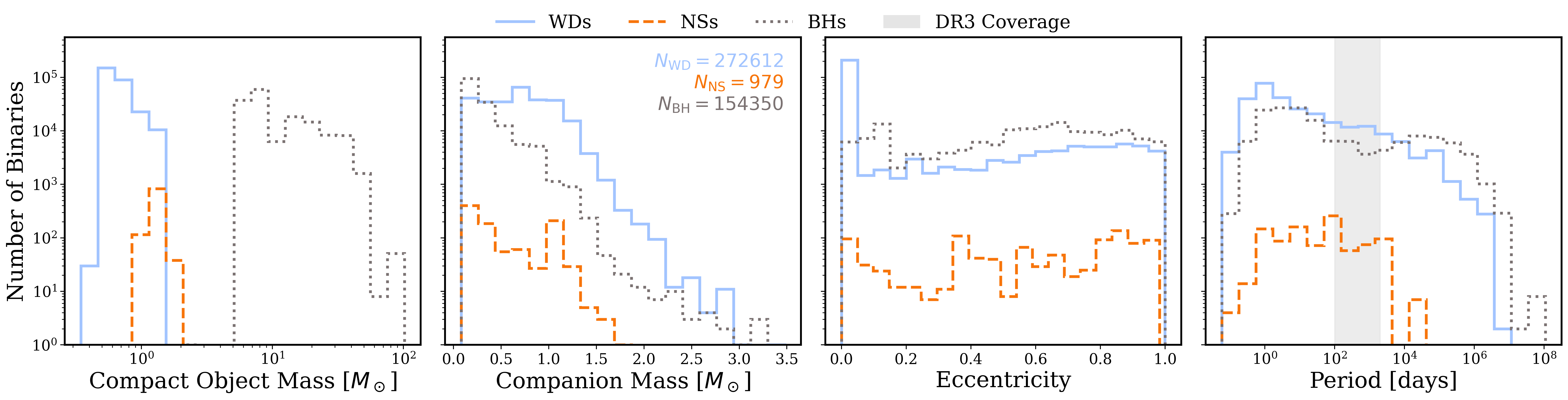}
    \caption{The properties of COs in binaries with luminous companions released to the Galactic field from disrupted dense clusters. Most of these binaries contain a WD, followed by BHs, while NSs are the least numerous. The WDs have comparatively shorter periods and orbits that are more circular compared to the other COs. The dynamically formed binaries (all NS and BH binaries in this study) extend to low eccentricities, showcasing that low eccentricity is not necessarily indicative of a lack of dynamics. The BHs also favour lower mass companions compared to the WDs, making them more difficult to observe.}
    \label{fig:cmc_histogram}
\end{figure*}

As discussed and shown in Figs. \ref{fig:emosaic_disruptiontime_mass} and \ref{fig:emosaic_properties}, the disrupted massive clusters in EMP-\textit{Pathfinder} have continuous distributions in initial masses, metallicities, and galactocentric radii. In contrast, the \texttt{CMC} models consist of grids in these properties as explained in Section \ref{sec:cmc}. To bridge the EMP-\textit{Pathfinder} simulated clusters to the \texttt{CMC} models, for each disrupted cluster from EMP-\textit{Pathfinder} we select the closest model from the \texttt{CMC} suite using two mapping techniques. First, we map only by the initial cluster masses: the `Mass' map. The other properties, metallicity and virial radius, are drawn randomly from the relevant values for the \texttt{CMC} catalog models. Second, we additionally account for cluster metallicities: the `Mass + Metal' map. The EMP-\textit{Pathfinder} metallicities (Figure~\ref{fig:emosaic_disruptiontime_mass}) are grouped into three bins, $Z \leq 0.00065$, $0.00065 < Z \leq 0.0065$, and $Z > 0.0065$, centered on the three metallicities of the \texttt{CMC} catalog models. We then extract CO binaries from the selected \texttt{CMC} clusters at approximately the cluster disruption time inferred from EMP-\textit{Pathfinder}(Figure~\ref{fig:emosaic_disruptiontime_mass}). We also track binaries that were ejected from the clusters prior to the disruption. Since the EMP-\texttt{Pathfinder} cluster models have no set virial radius, for each disrupted cluster matched in mass and metallicity, we randomly draw an initial virial radius for the \texttt{CMC} catalog models. The extracted CO-containing binaries are then evolved to the present day from the time of cluster disruption with the \texttt{COSMIC} binary population synthesis package \citep{COSMIC}. Furthermore, due to mass segregation, most CO binaries are concentrated in the cluster cores; we assume they are not lost during the gradual tidal stripping of their host clusters' outermost regions, remaining instead at the location of final cluster disruption.

The distributions of CO-containing binary systems sampled from varied \texttt{CMC}/EMP-\textit{Pathfinder} mapping strategies are displayed in Fig. \ref{fig:histograms_mappings}. We observe negligible stochastic variation between samples generated using different random draws. The differences induced by distinct mapping criteria (including metallicity vs. not) are minimal, where the `Mass + Metal' map yields fewer massive BH ($>10 M_\odot$) binaries compared to the `Mass' map by $\approx 29\%$. This reduction occurs because the metallicity constraint forces more binaries to be drawn from high metallicity clusters; $\approx 52\%$ of EMP-\textit{Pathfinder} disrupted clusters have metallicities above $[Fe/H]\geq-0.3=1/2 Z_\odot$. As a consequence, there are fewer massive BHs, which form more efficiently at low metallicity. Since these massive remnants are subject to more frequent dynamical processing, their depletion suppresses the high-eccentricity and long-period tails of the distribution. For the remainder of this work, we adopt the first random draw version of the `Mass + Metal Map' as this is the more physically motivated mapping, accounting for cluster metallicity.

\subsection{Binaries From Disrupted Clusters}\label{subsec:binaries_disrupted}

Here we present the intrinsic properties of the CO--luminous companion binaries that are released from disrupted dense star clusters in our hybrid \texttt{CMC}/EMP-\textit{Pathfinder} model. In total, approximately $4\times10^5$ COs in binaries with luminous companions were released to the field via cluster disruption. Fig. \ref{fig:cmc_histogram} shows the properties of these binaries, separated by CO type. WDs are the most numerous at $N_\text{WD} \approx 3\times10^5$, followed by $N_\text{BH}\approx1.5\times10^5$, and finally $N_\text{NS}\approx 1\times10^3$. Most binaries contain a WD as a result of the initial mass function, which causes less massive objects to be more abundant. This would also cause NS binaries to be the second most common. However, in dense cluster environments, BHs dominate dynamical interactions in the cluster cores due to mass segregation. As a result, many BH--luminous star binaries can form even though BHs themselves are less common. At the same time, many NSs are ejected from their host clusters by supernova natal kicks. Moreover, BH burning further suppresses NS mass segregation toward the cluster core, reducing their participation in dynamical interactions that would otherwise form binaries. Consequently, only core-collapsed star clusters are able to efficiently form large numbers of NS binaries \citep{Ye+2019_msp}. Thus, clusters are efficient producers of binaries containing BHs and WDs. 

\begin{figure*}
    \centering
    \includegraphics[width=1\linewidth]{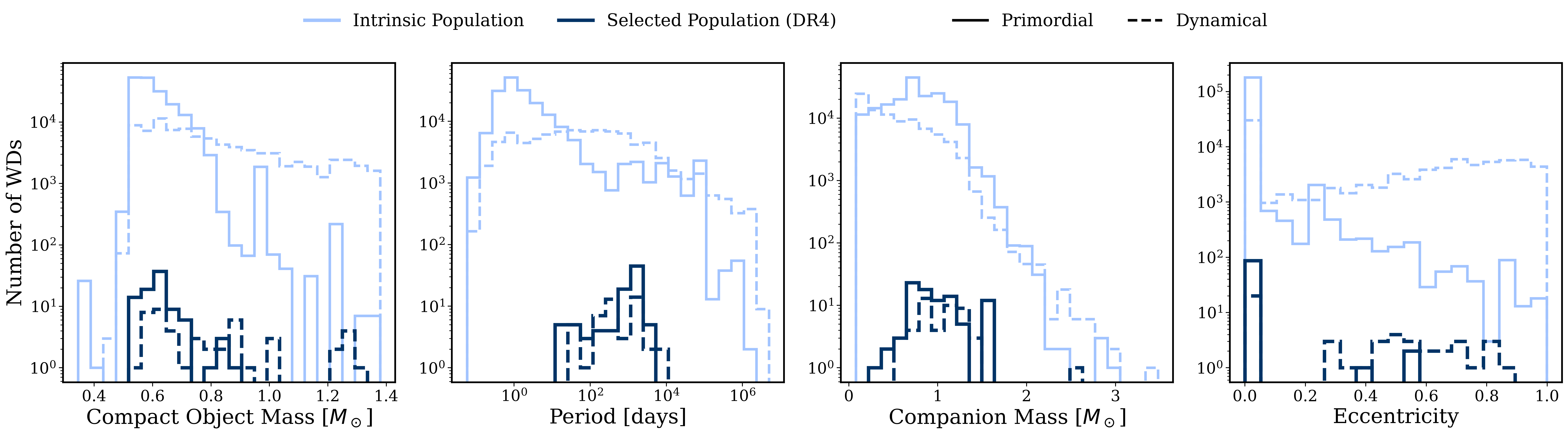}
    \caption{Property distributions of primordial (solid) and dynamically-assembled (dashed) WD binaries with luminous companions released from disrupted clusters. Light blue indicates all of the samples from the CMC/EMP-\textit{Pathfinder} model, while dark blue indicates the systems that were `observable' at least once after passing through the Gaia selection function. This shows observations for DR4 over all realizations; the DR3 observations are encompassed by DR4 so not explicitly shown. The observed population selects against the low-period peak in the primordial binaries. The observations also reduce the number of binaries with low-mass companions.}
    \label{fig:wd_primordial_vs_dynamical}
\end{figure*}

Not only are the numbers of the different CO binaries distinct, so are the structures of their distributions in Fig. \ref{fig:cmc_histogram}. BH and NS binaries have a comparatively higher proportion of low-mass companions than WD binaries. As a consequence, WDs are more detectable than BHs even though they are similar in number; BH systems are fainter and therefore have less precise astrometry (the impact on detectability is analyzed in Section \ref{sec:detect}). This disparity in companion masses is caused by the predominant dynamical formation of BH and NS binaries that sample companions from the initial mass function, which has more low-mass main-sequence stars than higher-mass stars. In contrast, $\approx70\%$ of the WD binaries are primordial, with their companion masses and orbital properties shaped by binary evolution instead. Additionally, the WD binaries peak at low eccentricities and show more support at shorter orbital periods due to binary evolution and orbital circularization. Figure~\ref{fig:wd_primordial_vs_dynamical} more clearly illustrates the differences in the property distributions between primordial binaries and dynamically assembled binaries. Previous studies (e.g., \cite{2024OJAp....7E..27E}) have suggested that the low eccentricities of observed systems indicate a preference for isolated formation channels. However, we emphasize that dynamically assembled binaries in our model, including NS and BH binaries in Fig.~\ref{fig:cmc_histogram}, as well as dynamically assembled WD binaries in Fig.~\ref{fig:wd_primordial_vs_dynamical}, exhibit eccentricity distributions that extend to $e=0$.

\section{Detectability of Compact Object Binaries}\label{sec:detect}

We model the detectability of Gaia-like binaries from disrupted dense star clusters. We apply preliminary, non-stringent cuts to our dataset in Section \ref{subsec:gaiasens}, motivated by Gaia sensitivity. This process reduces the computational expense when running the binaries through the detailed Gaia selection pipeline to obtain orbital solutions, which we do in Section \ref{subsec:gaiamock}.

\subsection{Gaia Sensitivity}\label{subsec:gaiasens}

As mentioned in Section \ref{sec:COs}, we have predicted that approximately $4\times10^5$ CO--luminous companion binaries are released to the Galactic field from dissolved star clusters, and now we determine their detectability by considering the selection function of Gaia. Observational techniques capture particular systems more reliably than others, which imparts a selection function that biases the binaries observed from the true astrophysical population. We model these observational effects by imposing constraints on orbital period, distance, and apparent magnitude, as elaborated upon in the subsequent text.  

Gaia observations select for binaries with periods that are neither too short, making orbits unidentifiable via astrometry, nor too long, ensuring sufficient observational cadence to detect an orbit. We truncate our dataset with upper bounds of $P\approx 4000$ days for DR3 and $P\approx 7000$ days for DR4 \citep{2025PASP..137d4202N}. These limits are very inclusive; only a minority of systems within this orbital range will get an orbital solution. Additionally, we exclude binaries with $P<10$ days to remove those too close to be detectable by astrometry \citep{2023A&A...674A...9H}.

The EMP-\textit{Pathfinder} clusters have designated galactic locations (see Fig. \ref{fig:emosaic_coords}); therefore, to assess the uncertainties associated with the observational vantage point, we randomise the position of the Sun within the Galaxy over 100 realizations. We assume the Solar galactocentric distance is drawn from a normal distribution $\mathcal{N}(8.24, 0.2)$~kpc \citep{2023AstL...49..493B}, the azimuthal angle is drawn from a uniform distribution $U(0,2\pi)$, and the height above the galactic disk is set to 0.028~kpc \citep{2019MNRAS.482.1417B}. We then use the Galactocentric coordinates, $\textbf{r} = (x,y,z)$, of the clusters to calculate their heliocentric distance $d=|\textbf{r} - \textbf{r}_\odot|$. Gaia can only resolve binaries that are near the Sun, and we use the cutoffs of $<3$ kpc for DR3, and $<10$ kpc for DR4 to make cuts in our dataset.  

Additionally, we specify cuts based on apparent G-band magnitude, requiring $m_{G}<21$ \citep{2021A&A...649A...1G}. The absolute G-band magnitude, $M_G$, is obtained for each luminous companion using the \texttt{isochrones} package \citep{isochrones,Montet+2015}. Specifically, we utilized the MIST (MESA Isochrones and Stellar Tracks) stellar evolution models \citep{2016ApJ...823..102C}. The package employs an interpolation scheme within the three-dimensional parameter space of mass-age-[Fe/H] to estimate stellar properties, such as $M_G$, for values that lie between the pre-computed model grid points. For each of our binaries, we supplied their mass, age, and metallicity ($[Fe/H]$) derived from the CMC simulations to interpolate for their $M_G$.\footnote{For about 0.1\% of binaries, those with companion masses $> 1\,\text{M}_\odot$, $M_G$ cannot be determined because these systems fall outside the coverage of the interpolator grid in mass-age-[Fe/H] space.} We also apply dust extinction corrections to the magnitudes using \texttt{MWdust}, a package for 3D dust maps of the Milky Way \citep{2016ApJ...818..130B}. In particular, we use the \texttt{combined19} dust maps to obtain $E(B-V)$ \citep{2003A&A...409..205D, 2006A&A...453..635M, 2019ApJ...887...93G}. 

We compute the extinction using $A_G = 2.8 E(B-V)$, similar to the calculation in \citet{2025PASP..137d4202N}. Finally, the G-band magnitude is calculated as:
\begin{equation}
    m_G = M_G + 5 \log_{10}\left(\frac{d}{10 \text{pc}}\right) + A_G
\end{equation}
for which we discard any systems with $m_G > 21$.

Following these selection cuts, each remaining binary was assigned random orbital orientations (inclination $i$, longitude of ascending node ${\Omega}$, argument of periastron $\omega$) and orbital phase, $T_p$. We remove systems with photocenter semi-major axes $a_0<0.1$ mas, below which the astrometric wiggle of a binary would be too minuscule to detect \citep{2025PASP..137d4202N}. 

The results of applying these initial selection cuts on our dataset are summarized in Table \ref{tab:selection}. In DR3, distance accounts for the largest cuts, with $\approx 99\%$ of binaries discarded on this attribute alone. The final number of binaries that pass all preliminary cuts is only 0.02\% of the initial sample, a sparse 41--127 (90\% credible range) binaries eligible for detection, depending on the randomized location of the Sun). The majority of these are binaries containing WDs. In contrast, the small initial number of NS binaries is reduced down to 0-2 systems. The BH binaries also undergo a substantial reduction, down to 0-11 systems, $<0.01\%$ of their initial number. 

On the other hand, the gain in observing volumes in DR4 allows 1624--2593 binaries to pass the selection cuts, expanding the eligible number of binaries by a factor of $\approx 30$ compared to DR3. The distance cut is no longer the bottleneck; only $\approx 33\%$ of binaries are at distances beyond the detection limit. Instead, the largest reduction of binaries in DR4 comes from astrometric precision and photometric constraints, which exclude $\approx 94\%$ and $\approx 87\%$ of the initial sample, respectively. In other words, the gain in observing volume is partially counteracted by sources appearing dimmer at greater distances. WD binaries dominate the demographics, accounting for $\approx 96\%$ of the final sample. BH binaries are suppressed despite comprising a significant fraction of the underlying population; $> 98\%$ are excluded by photometry alone. These BH binaries have lower mass companions than WDs, and therefore lower magnitude, making them more susceptible to magnitude truncation, as shown in Fig. \ref{fig:cmc_histogram}. This further demonstrates that the observable population is a skewed representation of the inherent astrophysical distribution and the formation channels and progenitor populations underlying it.

\begin{table}
\centering
\begin{tabular}{l|l|rrrrr} 
\toprule
 & CO & \multicolumn{5}{c}{\% Excluded} \\
\cmidrule(lr){2-2} \cmidrule(lr){3-7} \cmidrule(lr){7-7}
 & & Period & Dist. & $m_G$ & $a_0$ & \textbf{Total} \\
\midrule
\midrule
\multirow{4}{*}{DR3}
 & All & 71 & 99 & 87 & 94 & 0.02 \\
 & WD  & 73 & 99 & 80 & 92 & 0.03 \\
 & NS  & 36 & 99 & 85 & 93 & 0.1 \\
 & BH  & 68 & 99 & 98 & 97 & $>$0.01 \\
\midrule
\multirow{4}{*}{DR4}
 & All & 70 & 33 & 87 & 94 & 0.5\\
 & WD  & 72 & 31 & 81 & 92 & 0.8 \\
 & NS  & 36 & 25 & 86 & 93 & 0.8  \\
 & BH  & 66 & 36 & 98 & 97 & 0.05  \\
\bottomrule
\end{tabular}

\caption{The impact of the initial selection cuts on the sample of binaries from disrupted dense star clusters. We report the median percentages after computing the eligible sample from 100 realizations of the Solar System's position in the Galaxy. Note that the reduction in the sample by each selection criterion is calculated independently of the others. In DR3, the largest cuts come from the distance selection, but in DR4, the largest cuts are from the requirements for apparent magnitude and photocenter semi-major axis. These values do not reflect the final detection yields; the final counts of detected binaries are detailed in Table \ref{tab:detections}.}
\label{tab:selection}
\end{table}

The binaries that pass these initial cuts are still unlikely to actually be detected. We run these eligible systems through \texttt{gaiamock} to fully model the Gaia selection function in Section \ref{subsec:gaiamock}.

\subsection{Gaia Selection Function}\label{subsec:gaiamock}
Systems that satisfy the aforementioned selection criteria and are eligible for detection are subsequently passed to \texttt{gaiamock} \citep{2024OJAp....7E.100E}, which simulates the full Gaia astrometric pipeline. Gaia detects binaries by running sources through a `cascade' that attempts to determine which model best explains the astrometric signal, progressing from models of lower complexity to those of higher complexity \citep{2023A&A...674A...9H, 2024NewAR..9801694E, gaiamock}. The pipeline identifies sources that are poorly fit to the 5-parameter single star model, and progresses to a 9-parameter (variable acceleration) and 7-parameter (constant acceleration) fit. If these models fail to fit the data to a required significance, the full orbital solution is computed. The detections that pass through the entire cascade can be spurious, so a set of empirical quality cuts is then applied (see Section \ref{sec:detect} and \citet{2023A&A...674A...9H}). We utilize the \texttt{gaiamock} function \texttt{run\_full\_astrometric\_cascade} to model the observations of our eligible binaries. A binary counts as detected if it satisfies criteria for signal-to-noise ratios and orbital characterization thresholds. For both DR3 and DR4, we require the signal-to-noise ratio of the photocenter semi-major axis to comply with $\chi_{a_0} = a_0  /\sigma_{a_0}>5$. In addition, we require the signal-to-noise ratio of the parallax to be $\chi_\varpi = \varpi / \sigma_\varpi > 5$ in DR4. To compute the selection functions of DR3, we require the more stringent additional criteria $\chi_\varpi > \frac{20,000}{P}$, as well as $\chi_{a_0}>\frac{158}{\sqrt{P}}$ and $\sigma_\text{ecc}<0.079 \ln{P}-0.244$. Details on the \texttt{gaiamock} pipeline and these detection criteria can be found in \citet{2023A&A...674A...9H} and \citet{2024OJAp....7E.100E}.

\subsection{Detectable Binaries from Disrupted Clusters}\label{subsec:detectable_binaries}

\begin{figure*}[h!]
    \centering
    \includegraphics[width=1\linewidth]{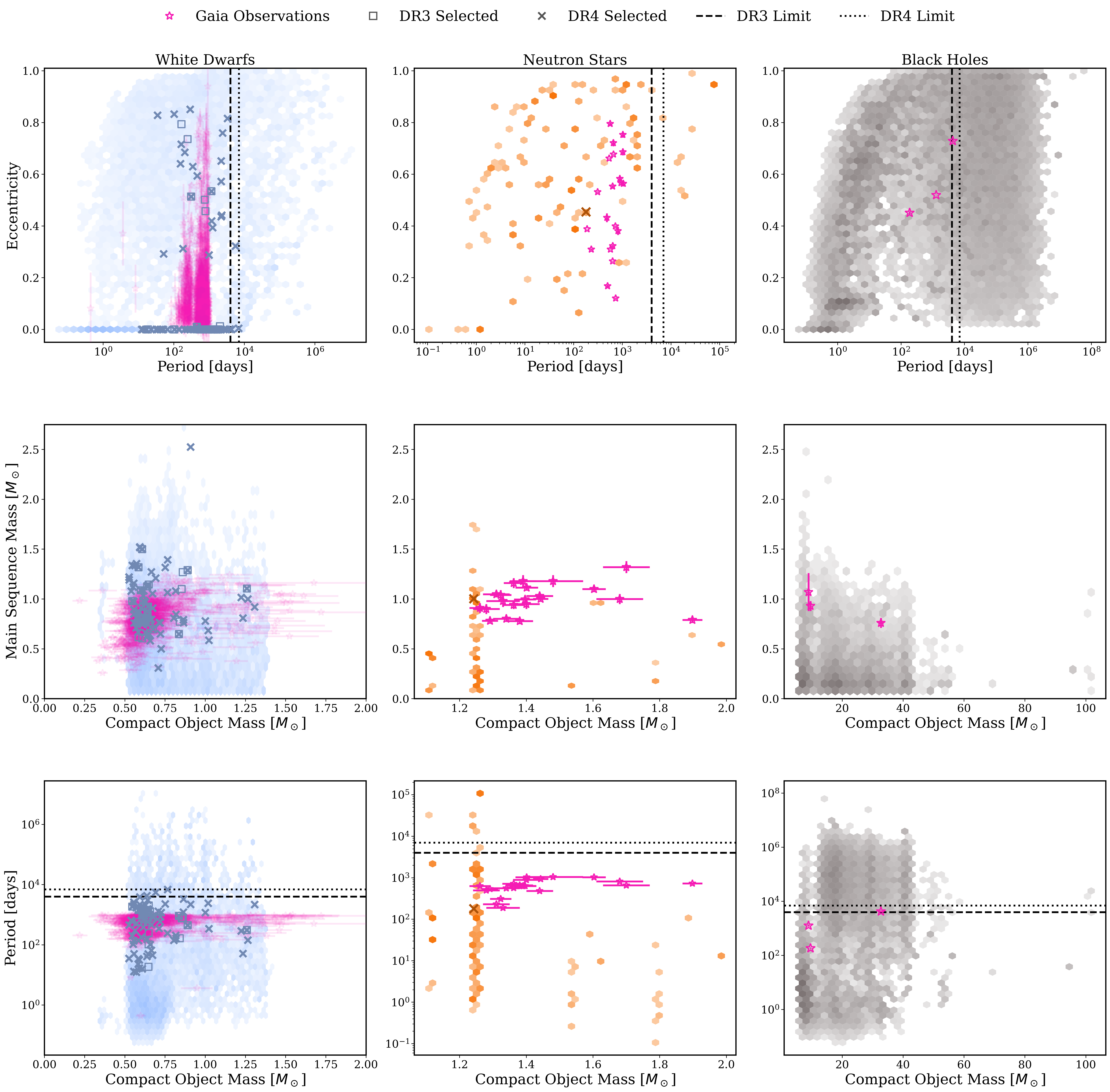}
    \caption{Properties of the CMC/EMP-\textit{Pathfinder} binaries. Blue shading represent WDs, orange represent NSs, and gray represents BHs. The pink stars indicate Gaia detections. We display the WD candidates as selected by \citet{2024MNRAS.529.3729S}. Squares indicate detections over all 100 realizations; filled symbols represent DR3 and open symbols represent DR4. Note that this visualization displays a factor of 100 more detections than would be expected in any single realization, as it represents the total marginalized discovery space. The dashed (dotted) line shows the maximum period detectable in Gaia DR3 (DR4). 
    \textit{Top:} Eccentricity vs. period. The observable points clearly showcase the orbital period selection. Approximately 66\% of BHs in the model are in systems too wide to pass our initial lenient DR3 and DR4 cuts.
    \textit{Middle:} Luminous companion mass vs. CO mass. Our model produced a smaller proportion of massive NSs due to the adopted supernova prescription (see Section \ref{sec:cmc}).
    \textit{Bottom:} Period vs. CO mass.}
    \label{fig:cmc_properties_detected}
\end{figure*}

We overplot the detectable binaries (in both Gaia DR3 and DR4) from the \texttt{CMC}/EMP-\textit{Pathfinder} model on top of the intrinsic binary populations in Figure~\ref{fig:cmc_properties_detected}using the selection functions discussed above, and compare these detectable binaries to the observed Gaia CO binaries. The top panels of Fig. \ref{fig:cmc_properties_detected} show the eccentricity vs. period parameter space for the CO-luminous companion binaries. The Gaia observations overlap with the coverage of our model binaries. The observed WD binaries exhibit eccentricities spanning the full range from $e=0$ to $e \approx 1$. The periods of the detectable binaries are concentrated within the range dictated by the selection function. Most of the model BHs have periods beyond our upper limits for the Gaia detection horizon of 4000 days in DR3 ($\approx 68\%$) and 7000 days in DR4 ($\approx 66\%$; see Table \ref{tab:selection}). This indicates that the majority of these modeled binaries are unobservable in these data releases. Furthermore, even among the systems that pass the initial cuts in Section \ref{subsec:gaiasens} and are processed through the \texttt{gaiamock} pipeline, a large fraction are still excluded due to insufficient orbital coverage. Relative to the Gaia observational baseline \citep[66 months for DR4;][]{2024A&A...686L...2G}, the orbital coverage of BHs (median 0.3 orbits) is significantly lower than for WDs (1.1) or NSs (1.0). Correspondingly, 87\% of modeled BHs fail to sample even half an orbital period, whereas only 23\% of WDs and $<0.1$\% of NSs fall below this threshold. Thus, despite having similar intrinsic abundance to WDs, BHs are not detectable in our model due to their characteristically long orbital periods.

Also shown in Fig. \ref{fig:cmc_properties_detected} are the CO and main sequence masses. The observed masses of the model WDs and their companions broadly cover their underlying mass ranges, although there is a higher density of observations at the lower end of the mass ranges due to the bottom-heavy nature of the initial mass function.

In DR3, the number of detectable binaries originating from disrupted clusters is sparse, as summarized in Table~\ref{tab:detections}. We predict a mean of only $0.3$ observable WDs (90\% credible range: 0--2) over 100 realizations; whereas there are nearly 3200 WD containing binaries observed so far \citet{2024MNRAS.529.3729S}. As expected from the initial population and observational selection cuts, the prospects for NS binaries are similarly limited. The expected yield is 0 to 90\% credibility. Thus, our model suggests disrupted dense star clusters do not account for the over-representation of metal-poor halo NS binaries in Gaia DR3. Likewise, BH binaries are virtually undetectable with 0 detections. As evident in Table \ref{tab:selection}, the lack of detections is a consequence of most binaries from disrupted dense star clusters being too distant for DR3 sensitivity.

The outlook improves slightly in the expansion to DR4. The model anticipates a mean of $5.5$ WDs (90\% credible range: 1--14), representing an approximately ten-fold improvement over the DR3 prediction. The NS average also trends upward to just $0.02$; detections remain rare events with the credible interval remaining [0,0]. This lack of significant improvement is driven by several physical bottlenecks. As shown in Table~\ref{tab:selection}, the most significant constraints on BH detection in DR4 are photometry and astrometry, largely because these BHs are paired with comparatively low-mass stellar companions (see Fig. \ref{fig:cmc_histogram}). Furthermore, the BH population is characterized by long orbital periods (Figs. \ref{fig:cmc_histogram} and \ref{fig:cmc_properties_detected}), the majority of which exceed the temporal baseline of Gaia observations. Consequently, the survey cannot sample enough of the orbits to produce a robust astrometric solution for the bulk of the modeled BH population.

\begin{table}
\centering
\begin{tabular}{l|l|r|r}
\toprule
Release & CO & Mean & 90\% CI \\
\midrule
\midrule
\multirow{3}{*}{DR3} 
 & WD & 0.33 & [0, 2] \\
 & NS & 0.00 & [0, 0] \\
 & BH & 0.00 & [0, 0] \\
\midrule
\multirow{3}{*}{DR4} 
 & WD & 5.45 & [1, 14] \\
 & NS & 0.02 & [0, 0] \\
 & BH & 0.00 & [0, 0] \\
\bottomrule
\end{tabular}
\caption{Detection statistics for the CO--luminous binary populations. Values represent the number of detected systems in a single realization, averaged over 100 realizations.}
\label{tab:detections}
\end{table}

\begin{figure}[h!]
    \centering
    \includegraphics[width=1\linewidth]{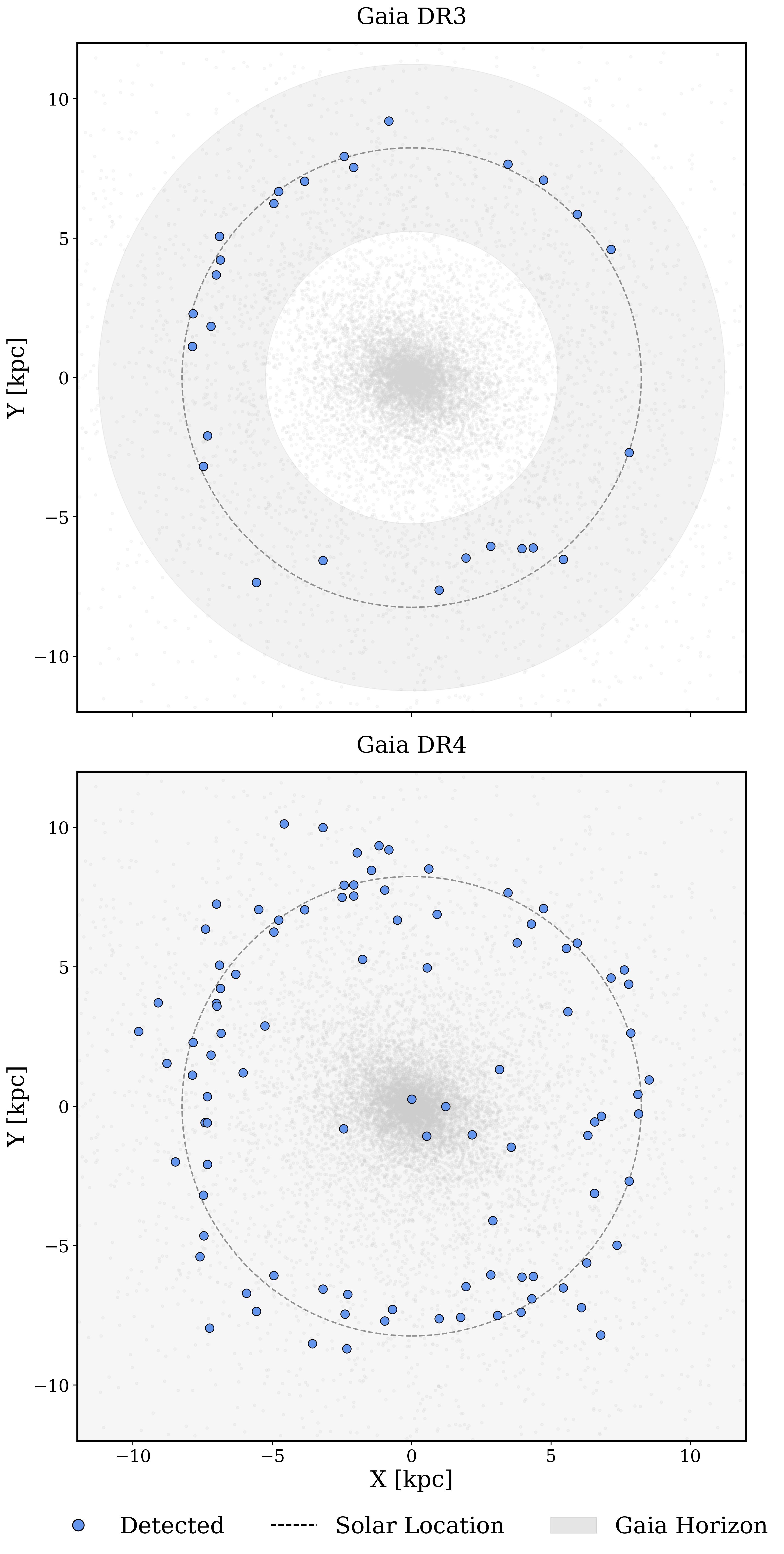}
    \caption{The spatial locations of our model binaries, based on the EMP-\textit{Pathfinder} cluster locations in Fig. \ref{fig:emosaic_coords}, in cartesian coordinates. The binaries detected at least once across the 100 realizations are shown as blue circles.The Solar location is marked by the dashed line. The detection horizon of DR3 (top) and DR4 (bottom) is shown by the grey shaded region. The DR4 horizon fully covers the plot.}
    \label{fig:emosaic_coordinates_detected}
\end{figure}

Fig. \ref{fig:emosaic_coordinates_detected} shows the locations of the model binaries that were detected at least once throughout the 100 realizations of our simulated observations. Most detectable binaries are located within the Solar neighbourhood, but most clusters dissolved closer to the Galactic centre, too distant to be within Gaia's detection horizon in DR3. These binaries are technically within the DR4 horizon. However, the number of additional detections is stifled by the decrease in apparent brightness inherent to sources located in the Galactic centre from increased distance and dust.

\section{Discussion}\label{sec:discuss}
\subsection{Model Uncertainties}\label{subsec:uncer}
Assumptions in modeling cluster formation and disruption, as well as in stellar and binary evolution in dynamical environments, introduce uncertainties in the predicted number and properties of CO binaries originating from disrupted star clusters.

A primary source of uncertainty in this analysis is the assumption that binaries remain at the cluster's final coordinates at dissolution. In reality, the galactic orbits of these systems are shaped by their initial ejection velocities from the host clusters, followed by further orbital migration due to secular galactic evolution and external perturbations. In addition, the time resolution of the snapshot output in EMP-\textit{Pathfinder} introduces inherent uncertainty in the precise locations where clusters are disrupted. However, we have verified that this does not qualitatively impact our results. Within a 10~kpc radius of the Sun, the volume is large enough to include the high-density regions of the inner Galaxy. Since many clusters here reside on relatively stable orbits, the physical distance they travel between time steps remains limited. Conversely, the 3~kpc local radius around the Sun has a relatively lower cluster density, so many of the clusters within this volume originated from more populated regions elsewhere in our model. However, because clusters are just as likely to enter the 3~kpc volume as they are to leave it between snapshots, the net number of systems within this region, and thus the number of detectable binaries, is not expected to change significantly. This implies that our findings of a scarcity of detectable binaries is further supported. 

In addition to uncertainties for cluster disruption in the Galactic environment, there are also uncertainties from assumptions of supernova physics and binary star evolution in \texttt{CMC} star cluster models. For example, we assume a supernova natal kick distribution for NSs derived from observed isolated pulsars in the Galaxy \citep{Hobbs+2005}. However, more recent studies, including ones that also consider the observed populations of NSs in binary systems, suggest that the natal kick for NSs may be lower than previously thought \citep[e.g.,][]{ODoherty+2023,Disberg_Mandel_2025}. If this is the case, more NSs would be retained in dense star clusters and participate in subsequent dynamical interactions that may form Gaia-like NS binaries. We leave a detailed exploration of the impacts of NS natal kicks on CO binary formation for future studies.

Furthermore, the \texttt{CMC} catalog models adopt one set of initial binary properties (Section~\ref{sec:cmc}) and fix the treatments for all binary and stellar evolution processes, including those for mass transfer stability and common envelope evolution. It is well understood that stability in mass transfer can significantly alter the final binary orbital periods \citep[e.g.,][]{Soberman+1997}, which are essential for understanding the evolution of CO binaries detected by Gaia \citep[e.g.,][]{Rubio+2025}. We have verified that changing the initial binary orbital period and eccentricity distributions from a log-normal and thermal distribution, respectively, to distributions motivated by observations of Galactic O-type stars \citep{Sana+2012}, does not significantly influence the number or binary properties of CO--MS binaries produced in dense star clusters. This is likely because the log-normal orbital period distribution closely resembles that inferred by \citet{Sana+2012}, especially at large orbital periods, while subsequent binary evolution largely erases the imprint of the initial eccentricity distribution. 

\begin{figure}[h!]
    \centering
    \includegraphics[width=\columnwidth]{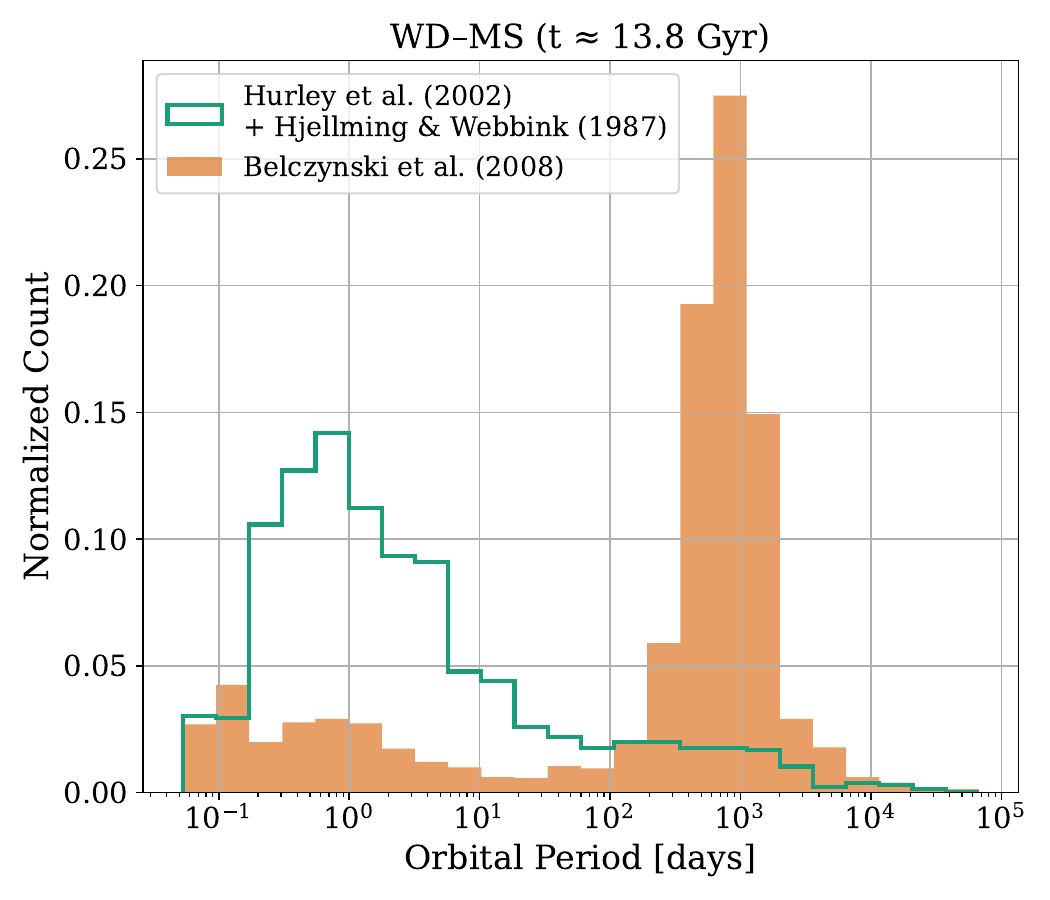}
    \caption{Orbital period distributions of model WD--MS binaries (primordial and dynamically assembled) from dense star clusters at the present day, comparing two binary mass transfer stability criteria. All other initial conditions and binary evolution physics are identical across simulations. Each distribution is normalized by the total number of WD--MS binaries over two \texttt{CMC} runs with different random seeds. The green histogram corresponds to models using the default \texttt{BSE} prescription for mass transfer stability \citep{Hurley+2002}, with asymptotic giant branch and giant branch stars treated following \citet{Hjellming&Webbink1987}. The orange histogram shows models adopting the prescription of \citet{Belczynski+2008} as an example.}
    \label{fig:masstransfer}
\end{figure}

On the other hand, we found that mass transfer stability can significantly affect the number and binary properties of Gaia-detectable CO--MS binaries originated from dense star clusters. For example, Figure~\ref{fig:masstransfer} shows the WD--MS binaries from dense star clusters simulations with two different mass transfer stability criteria. These example simulations all have the same initial conditions, with the initial number of stars $N=8\times10^5$, virial radius $r_v=1~$pc, galactocentric distance $r_g=8~$kpc, and metallicity $Z=0.002$. All other binary evolution treatments are fixed except for the mass transfer stability. The updated mass transfer prescription from \citet{Belczynski+2008} allows for more stable mass transfer and many more WD binaries to have wider orbital periods around 1000~days, potentially detectable by Gaia, while the older prescription from \citet{Hurley+2002} and \citet{Hjellming&Webbink1987} has fewer WD binaries at orbital periods within the Gaia detection limit. \citet{Yamaguchi+2025} studied the effects of mass transfer stability in isolated binaries on the population of \textit{Gaia} WD binaries and suggested that for accretor-to-donor mass ratios above $\sim 0.4$, mass transfer is stable, enabling the production of wide binaries. Similarly, \citet{Rubio+2025} found that more stable mass transfer during the first ascent giant branch phase in isolated binaries is favored by APOGEE-GALEX-Gaia observations. We leave a systematic exploration of the effects of mass transfer stability in dynamical environments to future work.

We have verified that our results are robust to the sampling of our model binaries. Using the `Mass' map introduced in Section \ref{sec:map} and Fig. \ref{fig:histograms_mappings} instead of the `Mass + Metal' map between EMP-\textit{Pathfinder} and \texttt{CMC} does not impact the results. Recall, the `Mass + Metal' map reduces the number of BHs by $\approx 29\%$. However, implementing the `Mass' map increased the mean number of BH detections from 0 to 0.03 in DR4, with no change in DR3. Lastly, the initial selection boundaries in Section \ref{subsec:gaiasens} were chosen to be lenient; applying stricter cuts would only further diminish our already modest detection yield. In contrast, extending the acceptable period range is offset by the insufficient orbital coverage available for long-period systems. Similarly, expanding the distance limits is counteracted by source faintness and the associated degradation in astrometric precision for distant binaries. Because our constraints on $m_G$ and $a_0$ are motivated by established Gaia Collaboration standards, we consider these parameters fixed for the purpose of this analysis.

\subsection{Potential Formation Pathways of Gaia Compact Object Binaries}\label{subsec:form_way}

Since their discovery, Gaia CO binaries have been particularly intriguing for studying the evolution of stellar remnants without contamination from their luminous companions. However, standard isolated binary evolution often struggles to reproduce these systems, due, for example, to uncertainties in common-envelope evolution. Alternative formation channels are proposed, including dynamical formation in star clusters of various masses \citep[e.g.,][]{Rastello+2023,DiCarlo+2024,MarinPina+2024,Tanikawa+2024}.

The discovery of Gaia BH3 in the ED-2 stream \citep{2024A&A...686L...2G, 2024A&A...687L...3B} provides strong evidence that at least a fraction of BH--MS binaries in the Galaxy originate from disrupted star clusters. For example, \citet{DiCarlo+2024} found that the dynamical formation of BH--MS binaries in open and young star clusters is more efficient at producing Gaia-like BH--star systems than isolated binary evolution, and may dominate the observable population even if only $\sim10\%$ of star formation occurs in clusters. Similarly, \citet{MarinPina+2024} and \citet{Rastello+2023} also suggested high formation efficiencies of BH--MS binaries in cluster environments. In the EMP-\textit{Pathfinder}/\texttt{CMC} model, we find that the total BH--MS formation efficiency is $\sim3\times10^{-5}\,M_{\odot}^{-1}$, broadly consistent with the values reported in \citet{MarinPina+2024} and \citet{Fantoccoli+2025}. 

However, the relative formation efficiency of BH--MS binaries originating from star clusters versus isolated binary evolution remains highly uncertain. \citet{Kotko+2024} suggested that both channels produce Gaia-like BH binaries with comparable efficiencies, $\sim3\times10^{-7}$ per unit stellar mass. \citet{2025PASP..137d4202N} used \texttt{gaiamock} to forward model the Gaia selection function on binary populations in a simulated Milky Way–like galaxy with a realistic, metallicity-dependent star formation history and 3D dust map, analyzing isolated binaries and open and young star clusters separately. They found that the isolated binary evolution model of \citet{Chawla+2022} predicts no BH detections in DR3, significantly underestimating the observed Gaia BH population. In contrast, they found that the dynamical model of \citet{DiCarlo+2024} overpredicts the number of BHs with DR3 orbital solutions by a factor of $\sim8$. The discrepancy between the predicted observable results of \citet{DiCarlo+2024} and this study may stem from the higher formation efficiency of lower-mass star clusters compared to their massive, dense counterparts. Furthermore, lower-mass, young and open clusters can dissolve closer to the Solar neighborhood, instead of relying on tidal shocks from high-density cold clouds to disrupt the bulk of more massive clusters. In addition, there is a pile-up of binaries with an orbital period of $4\times10^3$~d in \citet{DiCarlo+2024} arising from their assumptions regarding binary properties, allowing for more Gaia detectable binaries.

The formation efficiency of NS--MS binaries in cluster environments is significantly lower than that of BH--MS binaries. We find that the NS--MS formation efficiency from dense star clusters is $\sim2\times10^{-7}\,M_{\odot}^{-1}$ given the adopted supernova and natal kick prescriptions, which is about two orders of magnitude lower than the BH--MS formation efficiency. Consistent with this trend, \citet{Tanikawa+2024} reported formation efficiencies of $\sim10^{-6}$–$10^{-5}\,M_{\odot}^{-1}$ for Gaia-like BH binaries, compared to $\lesssim10^{-7}\,M_{\odot}^{-1}$ for Gaia-like NS binaries in open star clusters with masses of $200$–$2000\,M_{\odot}$.

If most observed Gaia NS binaries originate from isolated binary evolution, the apparent overabundance of metal-poor, halo NS systems likely reflects the Galaxy’s assembly history, e.g., accretion of metal-poor dwarf galaxies, rather than dynamical formation in star cluster environments. In this scenario, a corresponding excess of metal-poor MS binaries should also be present in the halo. The absence of such a trend would instead point to alternative explanations. These might include an intrinsically higher formation efficiency of NSs in metal-poor environments, systematically lower natal kicks for NSs, which can enhance their retention in clusters and boost the dynamical formation of NS--MS binaries, or a larger initial population of denser, more massive star clusters.

\section{Conclusions}\label{sec:conclu}

In this work, we present the first study of detectable CO-luminous binaries from disrupted GC progenitors. We employ a novel methodology that couples the cosmological formation and disruption of star clusters with resolved stellar and binary evolution from Monte Carlo $N$-body models. We use EMP-\textit{Pathfinder} to simulate the properties of the Milky Way's dense clusters from their formation to dissolution, and map these clusters by mass to \texttt{CMC} models for the internal stellar evolution and dynamics. Motivated by recent Gaia discoveries, we tracked the populations of luminous-CO binaries that originated within these clusters. We simulated Gaia observations of our model binaries with the \texttt{gaiamock} pipeline to test the detectability of our model binaries and determine if dense star clusters are a feasible environment of origin for the Gaia CO-containing binaries.

We find that within our model:

\begin{itemize}
    
    \item Approximately $4\times10^5$ CO-luminous companion binaries are released to the Galactic field from dissolved dense star clusters, as discussed in Section \ref{subsec:binaries_disrupted}.
    
    \item Most of the released binaries contain a WD ($3\times10^5$), followed by those with a BH ($1.5\times10^5$), and lastly a NS ($1\times10^3$), as detailed in Section \ref{subsec:binaries_disrupted} and Fig. \ref{fig:cmc_histogram}.

    \item Despite a similar abundance to WDs, BHs are more difficult to detect because their companion mass distribution favors lower-mass stars, which are intrinsically faint. Furthermore, the characteristically longer orbital periods of BH binaries often exceed the Gaia mission baseline, preventing their orbits from being fully resolved.

    \item $\approx68\%$ of luminous-WD binaries are primordial (Section \ref{subsec:binaries_disrupted} and Fig. \ref{fig:wd_primordial_vs_dynamical}), whereas 100\% of the luminous-NS or BH binaries released from our disrupted dense clusters are dynamically formed. As a result, there are many more short-period, circular WD binaries.

    \item $<1\%$ of clusters disrupt within the approximate Gaia DR3 detection horizon, and up to $\approx 70\%$ disrupt within the approximate horizon of Gaia DR4, as reflected in Fig. \ref{fig:emosaic_properties}. As such, Section \ref{sec:detect} and Table \ref{tab:selection} detail that the largest bottleneck to detection in DR3 is distance, and in DR4 it is astrometric and photometric precision.
    
    \item Gaia detections are scarce for these systems, as explained in Section \ref{subsec:detectable_binaries} and Table \ref{tab:detections}. The predicted yields for Gaia DR3 are at most 2 WD and 0 NS and BH binaries (90\% credibility). For DR4, the expanded volume increases these detections to a maximum of 14 WD, however the estimated detection of NSs and BHs remains at 0. While dense star clusters can explain some of Gaia's CO--luminous binaries, not all of the observations can be accounted for by this formation channel.
    
\end{itemize}

These results have significant implications for the interpretation of Gaia CO binaries. First, our model predicts that disrupted dense clusters are inefficient producers of luminous-NS binaries, and consequently, we predict nearly zero detections for binaries containing NSs across all scenarios. This strongly suggests that disrupted dense clusters cannot account for the over-representation of metal-poor, halo NS binaries observed in Gaia DR3. Instead, this could reflect the Galaxy's assembly history of low-metallicity systems during the epoch of star cluster formation, or an overestimation of supernova natal kicks in current models. 

Additionally, our results highlight an observational bias against dynamically formed BH--luminous companion binaries. Although BH binaries make up a substantial fraction of the underlying population analyzed here, they remain virtually undetectable in DR3 and DR4. This suppression is driven by a high density of systems with long orbital periods (Fig. \ref{fig:cmc_properties_detected}) and their tendency to pair with lower-mass, dimmer main-sequence companions compared to WDs (Fig. \ref{fig:cmc_histogram}). This indicates that the observed Gaia population may be a biased representation of the true astrophysical demographics, with a selection against wide, dynamically formed binaries containing BHs.

In summary, the Gaia binary observations remain an intriguing probe of both CO formation, binary evolution processes, and the Milky Way's history. Existing studies of either isolated evolution or dynamical formation in star clusters have not fully accounted for the observed populations of WDs, NSs, and BHs in binaries with luminous companions. This work contributes to this effort by demonstrating that, while dense star clusters are a viable formation channel for a subset of these systems, they are insufficient to explain the full observed population. Future data releases will continue to improve the observed dataset, providing more opportunities to constrain possible formation channels. Coupled with more detailed modeling that unifies cluster dynamics, binary evolution, and Galactic structure, these observations will enable a more complete understanding of the origin and demographics of CO binaries in the Milky Way. 

\section*{Acknowledgments}
We thank Katie Breivik and Maya Fishbach for helpful discussions. A.S.Z. acknowledges the support of the Natural Sciences and Engineering Research Council of Canada - Canada Graduate Scholarships - Doctoral (NSERC-CGS-D) program, Ontario Early Researcher Award Grant ER24-18-170, and NSERC grant RGPIN-2023-05511. C.S.Y. and M.R.-C. acknowledge support from the Natural Sciences and Engineering Research Council of Canada (NSERC) DIS-2022-568580, and C.S.Y. acknowledges support from the Alfred P. Sloan Foundation. M.R.-C. acknowledges funding from the Global Talent Junior Fellowship Programme at the Instituto Galego de Física de Altas Enerx\'ias (IGFAE), supported by grant CEX2023-001318-M (Mar\'ia de Maeztu Unit of Excellence and Agencia Española de Investigaci\'on / 10.13039/501100011033). This work has received financial support from the Xunta de Galicia (CIGUS Network of Research Centres) and the European Union through the Galicia Feder 2021-2027 Program. This research was supported in part by grant NSF PHY-2309135 to the Kavli Institute for Theoretical Physics (KITP).

\software{
    \texttt{CMC} \citep{CMC1},
    \texttt{COSMIC} \citep{COSMIC},
    \texttt{isochrones} \citep{isochrones}, 
    \texttt{MWdust} \citep{2016ApJ...818..130B},
    \texttt{gaiamock} \citep{gaiamock}, 
    NumPy \citep{harris2020array}, 
    pandas \citep{mckinney2010data, pandas2020},
    Matplotlib \citep{hunter2007matplotlib}, 
    and Astropy \citep{astropy2013, astropy2018, astropy2022}
}

\bibliography{main}{}
\bibliographystyle{aasjournal}

\end{document}